\documentclass[conference]{IEEEtran}\IEEEoverridecommandlockouts
\usepackage{etoolbox}
\usepackage{algorithmic}
\usepackage{algorithm}

\makeatletter
\patchcmd{\@makecaption}
  {\scshape}
  {}
  {}
  {}
\makeatletter
\patchcmd{\@makecaption}
  {\\}
  {.\ }
  {}
  {}
\makeatother
 \usepackage{amsmath,amssymb}
 \usepackage{subfigure}
 \usepackage{graphicx,graphics,color,psfrag}
 \usepackage{cite,balance}
 \usepackage{caption}
 \allowdisplaybreaks
 \usepackage{algorithm}
 \usepackage{accents}
 \usepackage{amsthm}
 \usepackage{bm}
 \usepackage{algorithmic}
 \usepackage[english]{babel}
 \usepackage{multirow}
 \usepackage{enumerate}
 \usepackage{cases}
 \usepackage{stfloats}
 \usepackage{dsfont}
 \usepackage{color,soul}
 \usepackage{amsfonts}
 \usepackage{cite,graphicx,amsmath,amssymb}
 \usepackage{subfigure}
 \usepackage{fancyhdr}
 \usepackage{hhline}
 \usepackage{graphicx,graphics}
 \usepackage{array,color}
 \usepackage{amsmath}
\usepackage{float}
\usepackage{amssymb}
\usepackage{amsmath}
\usepackage{amsthm}
\usepackage{amsfonts}
\usepackage{graphicx}

\usepackage{epstopdf}
\usepackage{cite}
\usepackage{amsmath,bm}
\usepackage{subfigure}
\usepackage{graphicx}
\usepackage{color}
\usepackage{graphicx}
\usepackage{calc}
\usepackage{caption}
\newtheorem{proposition}{\textbf{Proposition}}

\begin{document}

\title{CDDM: Channel Denoising Diffusion Models for Wireless Communications}
\author{Tong Wu, Zhiyong Chen, Dazhi He, Liang Qian, Yin Xu, Meixia Tao, Wenjun Zhang\\
		Cooperative Medianet Innovation Center, Shanghai Jiao Tong University, Shanghai, China\\
		Email: \{wu\_tong, zhiyongchen, hedazhi, lqian, xuyin, mxtao, zhangwenjun\}@sjtu.edu.cn}
\maketitle
\begin{abstract}
Diffusion models (DM) can gradually learn to remove noise, which have been widely used in artificial intelligence generated content (AIGC) in recent years. The property of DM for removing noise leads us to wonder whether DM can be applied to wireless communications to help the receiver eliminate the channel noise. To address this, we propose channel denoising diffusion models (CDDM) for wireless communications in this paper. CDDM can be applied as a new physical layer module after the channel equalization to learn the distribution of the channel input signal, and then utilizes this learned knowledge to remove the channel noise. We design corresponding training and sampling algorithms for the forward diffusion process and the reverse sampling process of CDDM. Moreover, we apply CDDM to a semantic communications system based on joint source-channel coding (JSCC). Experimental results demonstrate that CDDM can further reduce the mean square error (MSE) after minimum mean square error (MMSE) equalizer, and the joint CDDM and JSCC system achieves better performance than the JSCC system and the traditional JPEG2000 with low-density parity-check (LDPC) code approach.
\end{abstract}

\section{Introduction}
\begin{figure}[t]
  \begin{center}
    \includegraphics*[width=8cm]{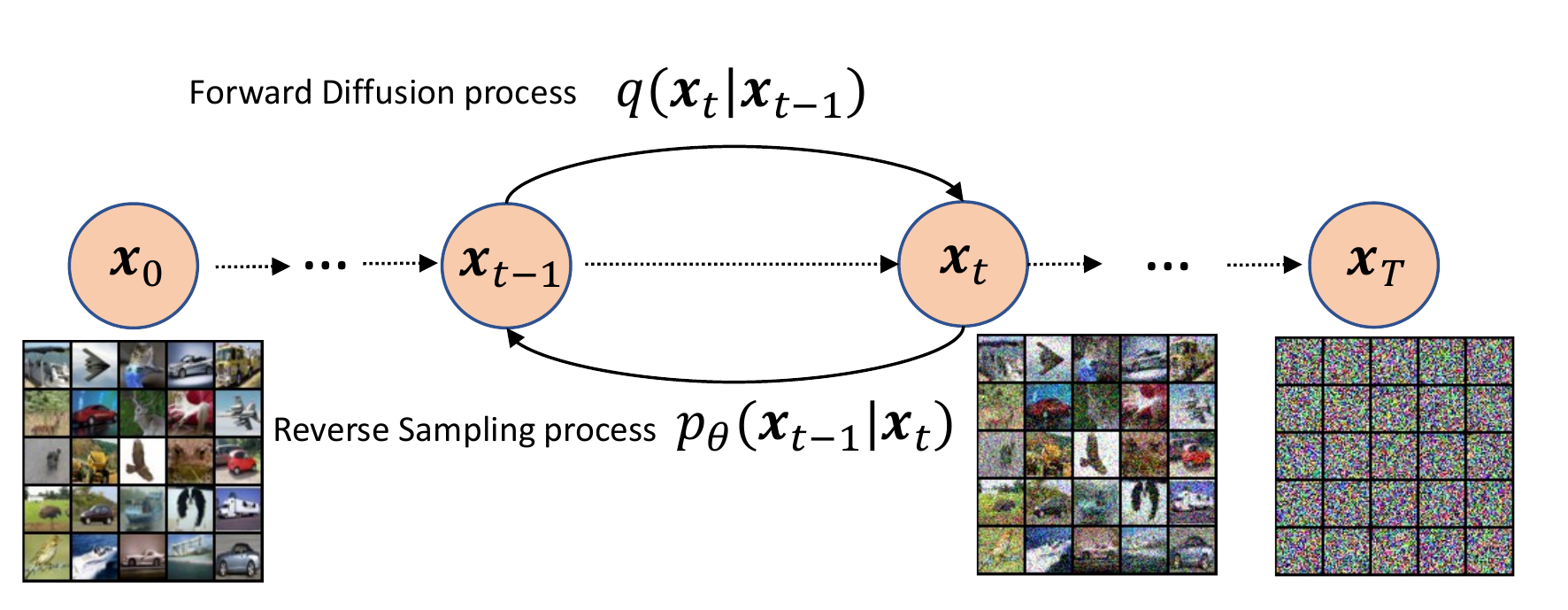}
  \end{center}
    \caption{\small{The forward diffusion process with transition kernel $q(\mathbf{x}_t|\mathbf{x}_{t-1})$ and the reverse sampling process with learnable transition kernel $p_\theta(\mathbf{x}_{t-1}|\mathbf{x}_t)$  of diffusion model in \cite{Ho}.}}
    \label{Diff}
\end{figure}

In machine learning, diffusion models (DM)\cite{DDPM2015, Ho,Song} have achieved unprecedented success in artificial intelligence generated content (AIGC) recently, including multimodal image generation and edition \cite{meng,Choi}, text, and video generation\cite{Lin,Sihyun}. DM gradually adds Gaussian noise to the available training data in the forward diffusion process until the data becomes all noise. Then, in the reverse sampling process, it learns to recover the data from the noise, as shown in Fig. \ref{Diff}. Generally, given a data distribution  $\mathbf{x}_0\sim q(\mathbf{x}_0)$, the forward diffusion process can generate the $t$-th sample of $\mathbf{x}_t$ by sampling a Gaussian vector $\epsilon \sim \mathcal{N}(0,\mathbf{I})$ as following
\begin{align}\label{DDPMforward}
  \mathbf{x}_t=\sqrt{\bar{\alpha}_t}\mathbf{x}_0+\sqrt{1-\bar{\alpha}_t}\epsilon,
\end{align}
where $\bar{\alpha}_t= {\textstyle \prod_{i=1}^{t}}\alpha_i $ and $\alpha_i\in(0,1)$ is a hyperparameter.

In wireless communications, it is well known that the received signal $y$ is a noisy and distorted version of the transmitted signal $x$, e.g., we have the following for the additive white Gaussian noise (AWGN) channel
\begin{align}
  y=x+n,
\end{align}
where $n$ is a white Gaussian noise.

Interestingly, compared to (1) and (2), we can find that the design approach of DM and wireless communications systems are similar. DM can gradually learn to remove noise, while the receiver in the wireless communications system is to recover the transmitted signal from the received signal. Clearly, \textbf{can DM be applied to the wireless communications system to help the receiver remove noise?} To the best of our knowledge, there have been no related works in the literature that address this question.

Motivated by this, we propose channel denoising diffusion models (CDDM) for wireless communications in this paper. CDDM can be applied as a new module after channel equalization to predict the channel noise and eliminate it, thereby enhancing the performance. We design the forward diffusion process based on the conditional distribution of the received signal after channel equalization (or without channel equalization) under Rayleigh fading channel (or AWGN channel). We design the corresponding training algorithm that solely relies on the forward diffusion process without any requirement of the received signal. The forward diffusion process also prompts us to design a sampling algorithm to achieve channel noise elimination.
%这里可以加一句加速采样的贡献

Furthermore, we apply the CDDM to a semantic communications system based on joint source-channel coding (JSCC) technique for wireless image transmission, where the signal after CDDM is fed into the JSCC decoder to recover the image. We test the mean square error (MSE) between the transmitted signal and the received signal after CDDM, and find that compared to the system without CDDM, the system with CDDM has smaller MSE performance both for Rayleigh fading channel and AWGN channel. This fact indicates that the proposed CDDM can effectively reduce the impact of channel noise through learning. The experimental results show that the joint CDDM and JSCC method outperforms both the JSCC method and the traditional JPEG2000 with low-density parity-check (LDPC) code approach in terms of the peak signal-to-noise ratio (PSNR) of the images.

\subsection{Related Works}
Compared to the prosperous researches of DM in AIGC, there are few works of DM in wireless communications so far. In \cite{kim}, DM is used to generate the wireless channel for a end-to-end communications system, and has almost the same performance as the channel-aware case. In \cite{Yoni}, DM with an adapted diffusion process is proposed for the decoding of algebraic block codes. 

In recent years, semantic communications \cite{Lan,jihong} have emerged as a new paradigmatic approach, characterized by its core idea of JSCC \cite{gundu2019, chenwei,XuTung,KeYang}, which considers the source and channel processes integrally based on deep neural network \cite{gundu2019}. Most studies on JSCC have designed specific JSCC frameworks for different data modals and achieved better performance compared with traditional wireless transmission schemes. In \cite{chenwei}, a novel JSCC method based on attention mechanisms is proposed, which can automatically adapt to various channel conditions. \cite{XuTung} introduces an adaptive deep learning based JSCC architecture for semantic communications. In \cite{KeYang}, the Swin Transformer \cite{Ze} is integrated into the deep JSCC framework to improve the performance of wireless image transmission. In summary, there have been no publications in literature regarding the joint design of DM and JSCC over wireless communications.

\section{Channel Denoising Diffusion Model}\label{I}
\begin{figure*}[t]
  \begin{center}
    \includegraphics*[width=15cm]{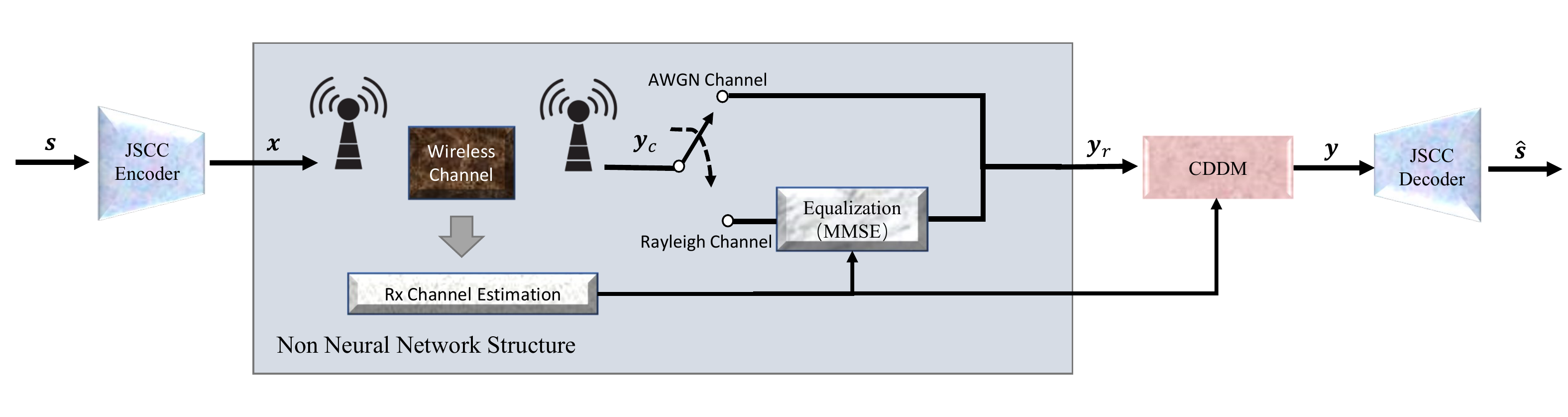}
  \end{center}
    \caption{\small{Architecture of the joint CDDM and JSCC system.}}
    \label{CDESC}
\end{figure*}

In this section, we describe the proposed CDDM which is placed after the channel equalization as shown in Fig. \ref{CDESC}. CDDM is trained using a specialized noise schedule adapted to the wireless channel, which enables it to effectively eliminate channel noise through a designed sampling algorithm.
\subsection{Conditional Distribution of the received signals}

Let $\mathbf{x}\in \mathbb{R}^{2k}$ be the real-valued symbols. Here, $k$ is the number of channel uses. $\mathbf{x_c}\in\mathbb{C}^{k}$ are the complex-valued symbols which can be transmitted through the wireless channel, and the $i$-th transmitted symbol of $\mathbf{x_c}$ can be expressed as ${x_{c,i}}={x_i}+jx_{i+k}$, for $i=1,...,k.$

Thus, the $i$-th received symbol of the received signal $\mathbf{y_c}$ is
\begin{align}\label{receive signal}
y_{c,i}=h_{c,i}x_{c,i}+n_{c,i}
\end{align}
where $h_{c,i}\in \mathbb{CN}(0,1)$ are independent and identically distributed (i.i.d.) Rayleigh fading gains, $x_{c,i}$ has a power constraint $\mathbb{E}[|x_{c,i}|^2]\leq 1$, and $n_{c,i}\in \mathbb{CN} (0,2\sigma^2)$ are i.i.d. AWGN samples.

%where $h_{c,i},n_{c,i}$ are the $i$-th elements of $\mathbf{h_c},\mathbf{n_c}$, $\mathbf{h_c}=\frac{1}{\sqrt{2}}(\mathbf{h}_1+j\mathbf{h}_2)$, $\mathbf{h}_1,\mathbf{h}_2 \sim \mathcal{N}(0,\mathbf{I}_k)$ for Rayleigh fading channel and $\mathbf{h}_1=\mathbf{h}_2=\mathbf{1}$ for AWGN channel. $\mathbf{n_c}=\frac{\sigma}{\sqrt{2}}(\mathbf{\epsilon}_1+j\mathbf{\epsilon}_2)$, $\mathbf{\epsilon}_1,\mathbf{\epsilon}_2 \sim \mathcal{N}(0,\mathbf{I}_k)$. Noise power is $\sigma^2$.

In this paper, we use minimum mean square error (MMSE) as an equalizer. $\mathbf{y_c}$ is then addressed by equalization as $\mathbf{y_{eq}}\in \mathbb{C}^{k}$, following a normalization-reshape module outputing a real vector $\mathbf{y_r}\in\mathbb{R}^{2k}$. We consider that the receiver can obtain the channel state $\mathbf{h_c}=[h_{c,1},...,h_{c,k}]$ through channel estimation. Therefore, we can have the conditional distribution of $\mathbf{y_r}$ with known $\mathbf{x}$ and $\mathbf{h_c}$, which can be formulated to instruct the forward diffusion and reverse sampling processes of CDDM.

\begin{proposition}\label{thm1}
With MMSE, the conditional distribution of $\mathbf{y_r}$ with known $\mathbf{x}$ and $\mathbf{h_c}$ under Rayleigh fading channel is
  \begin{gather}\label{zfyr}
    p(\mathbf{y_r}|\mathbf{x},\mathbf{h_c})\sim \mathcal{N}(\mathbf{y_r};\frac{1}{\sqrt{1+\sigma ^{2}}}\mathbf{W_s}\mathbf{x},\frac{\sigma ^{2}}{{1+\sigma ^{2}}}\mathbf{W}^2_{\mathbf{n}})
  \end{gather}
where $\mathbf{H_r}=diag({\mathbf{h_r}})$, $\mathbf{h}_{\mathbf{r}}=\begin{bmatrix}|\mathbf{h_c}|\\|\mathbf{h_c}|\end{bmatrix}\in \mathbb{R}^{2k}$, and
 \begin{gather}\label{WsMMSE}
   \mathbf{W_s}=\mathbf{H}^2_{\mathbf{r}}(\mathbf{H}^2_\mathbf{r}+2\sigma^2\mathbf{I})^{-1},\mathbf{W_n}=\mathbf{H_r}(\mathbf{H}^2_\mathbf{r}+2\sigma^2\mathbf{I})^{-1}.
 \end{gather}

\end{proposition}
\begin{IEEEproof}[Proof]
Based on the defination, $\mathbf{W_s}$ and $\mathbf{W_n}$ are diagonal matrix, where the $i$-th and ($i+k$)-th diagonal element are
  \begin{align}\label{Wselement}
    {W_{s,i}}= {W_{s,i+k}}= \frac{|h_{c,i}|^2}{|h_{c,i}|^2+2\sigma^2}\nonumber,\\
    {W_{n,i}}= {W_{n,i+k}}= \frac{|h_{c,i}|}{|h_{c,i}|^2+2\sigma^2}.
  \end{align}

The $i$-th output of MMSE ${y_{eq,i}}$ can be expressed as
  \begin{align}\label{zfeq}
    {y_{eq,i}}=\frac{|h_{c,i}|^2x_{c,i}+h_{c,i}^Hn_{c,i}}{|h_{c,i}|^2+2\sigma^2}.
  \end{align}

Based on (\ref{Wselement}), we have
  \begin{align}
    \frac{|h_{c,i}|^2x_{c,i}}{|h_{c,i}|^2+2\sigma^2}={W_{s,i}}x_{c,i}.
  \end{align}

With the resampling trick, the conditional distributions of real part and imaginary part of $\frac{h_{c,i}^Hn_{c,i}}{|h_{c,i}|^2+2\sigma^2}$ are
  \begin{align}\label{renoise}
    p(Re(\frac{h_{c,i}^Hn_{c,i}}{|h_{c,i}|^2+2\sigma^2})|h_{c,i}) &\sim\mathcal{N}(0,{\sigma^{2}}(\frac{|h_{c,i}|}{|h_{c,i}|^2+2\sigma^{2}})^2)\nonumber\\
&=\mathcal{N}(0,{\sigma^{2}}W_{n,i}^2),
  \end{align}
  \begin{align}\label{imnoise}
    %p(\sigma \frac{\epsilon_{2,i}*h_{1,i}-\epsilon_{1,i}*h_{2,i}}{2|h_{c,i}|^2}|h_{c,i})
    p(Im(\frac{h_{c,i}^Hn_{c,i}}{|h_{c,i}|^2+2\sigma^2})|h_{c,i})\sim \mathcal{N}(0,{\sigma^{2}}W_{n,i}^2).
  \end{align}
 
Accordingly, we can rewrite $\mathbf{y_r}$ as
  \begin{align}
    \mathbf{y_r}=\frac{1}{\sqrt{1+\sigma^{2}}}(\mathbf{W_sx}+\mathbf{n_r}),
  \end{align}
and the distribution $p(\mathbf{n_r}|\mathbf{h_c})$ is $\mathcal{N}(0,\sigma^{2}\mathbf{W}^2_{\mathbf{n}})$.

Therefore, we have 
\begin{gather}\label{zfyr_final}
  p(\mathbf{y_r}|\mathbf{x},\mathbf{h_c})\sim \mathcal{N}(\mathbf{y_r};\frac{1}{\sqrt{1+\sigma ^{2}}}\mathbf{W_s}\mathbf{x},\frac{\sigma ^{2}}{{1+\sigma ^{2}}}\mathbf{W}^2_{\mathbf{n}}).
\end{gather}
\end{IEEEproof}

Similarly, we have the following proposition for AWGN channel.
\begin{proposition}\label{thm2}
  Under AWGN channel, the conditional distribution of $\mathbf{y_r}$ with known $\mathbf{x}$ is
  \begin{align}
    p(\mathbf{y_r}|\mathbf{x})\sim \mathcal{N}(\mathbf{y_r};\frac{1}{\sqrt{1+\sigma ^{2}}}\mathbf{W_s}\mathbf{x},\frac{\sigma ^{2}}{{1+\sigma ^{2}}}\mathbf{W}^2_{\mathbf{n}})
  \end{align}
where $\mathbf{W_s}$ becomes $\mathbf{I}_{2k}$ and $\mathbf{W_n}$ becomes $\mathbf{I}_{2k}$ under AWGN channel.

\end{proposition}

Proposition 1 an Proposition 2 demonstrate that the channel noise after equalization and normalization-reshape can be re-sampled using $\mathbf{\epsilon} \sim \mathcal{N}(0,\mathbf{I}_{2k})$. Additionally, the noise coefficient matrix $\mathbf{W_n}$ is related to the modulo form of $\mathbf{h_c}$. As a result, $\mathbf{y_r}$ can be re-parametered as
\begin{align}
  \mathbf{y_r}=\frac{1}{\sqrt{1+\sigma ^{2}}}\mathbf{W_s}\mathbf{x}+\frac{\sigma}{\sqrt{1+\sigma ^{2}}}\mathbf{W_n}\epsilon.
\end{align}

\begin{figure*}[t]
  \begin{center}
    \includegraphics*[width=15cm]{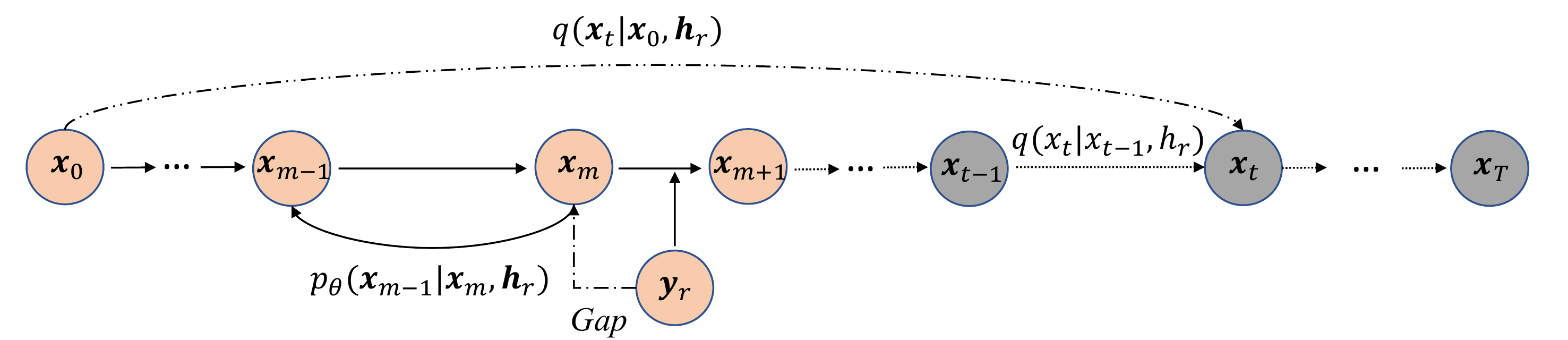}
  \end{center}
    \caption{\small{The forward diffusion process and reverse sampling process of the proposed CDDM.}}
    \label{ourstep}
\end{figure*}

Therefore, the proposed CDDM is trained to obtain $\mathbf{\epsilon_{\theta}}(\cdot )$, which is an estimation of $\mathbf{\epsilon}$. Here, $\mathbf{\theta}$ is model parameters. By using $\mathbf{\epsilon_\theta}(\cdot)$ and $\mathbf{W_n}$, a sampling algorithm is proposed to obtain $\mathbf{y}$ with the aim to recover $\mathbf{W_sx}$, which will be described in Section \ref{B}. The whole strcuture of the CDDM forward diffusion and reverse sampling process is illustrated in Fig. \ref{ourstep}.

\subsection{Training Algorithm of CDDM}
For the forward process of the proposed CDDM, the original source $\mathbf{x}_0$ is
\begin{align}
  \mathbf{x}_0=\mathbf{W_sx}.
\end{align}
Let $T$ be the hyperparameter. Similar to (1), for all $t\in \{1,2,...,T\}$, we define
\begin{align}\label{x_t}
  \mathbf{x}_t=\sqrt{\alpha _t}\mathbf{x}_{t-1}+\sqrt{1-\alpha _t}\mathbf{W_n}\mathbf{\epsilon},
\end{align}
and then it can be re-parametered as
\begin{align}\label{reforward}
  \mathbf{x}_t=\sqrt{\bar{\alpha}_t}\mathbf{x}_0+\sqrt{1-\bar{\alpha}_t}\mathbf{W_n}\mathbf{\epsilon}
\end{align}
such that the distribution $q(\mathbf{x}_t|\mathbf{x}_0,\mathbf{h_r})$ is
\begin{gather}\label{distributionforward}
  {q(\mathbf{x}_t|\mathbf{x}_0,\mathbf{h_r})\sim \mathcal{N}(\mathbf{x}_t;\sqrt{\bar{\alpha}_t}\mathbf{x}_0,({1-\bar{\alpha}_t})\mathbf{W}^2_{\mathbf{n}})}.
\end{gather}

Based on (\ref{zfyr}) and (\ref{distributionforward}), if $\bar{\alpha}_m=\frac{1}{1+\sigma^{2}}$, the Kullback-Leibler (KL) divergence is
\begin{align}
  D_{KL}(q(\mathbf{x}_m|\mathbf{x}_0,\mathbf{h_r})||p(\mathbf{y_r}|\mathbf{x}_0,\mathbf{h_c}))=0,
\end{align}
for $t=m$. This indicates that \textbf{CDDM can be trained on $\mathbf{x}_m$ instead of $\mathbf{y_r}$}. $\mathbf{x}_m$ is defined by $m$ steps as (\ref{x_t}) such 
that the predicted distribution by CDDM in reverse process can be decomposed into $m$ small steps and each of them is $p_{\mathbf{\theta}}(\mathbf{x}_{t-1}|\mathbf{x}_t,\mathbf{h_r})$ for $t\in \{1,2,...,m\}$.

The goal of CDDM is to recover $\mathbf{x}_0$ by learning the distribution of $\mathbf{x}_0$ and removing the channel noise. Therefore, the training of CDDM is performed by optimizing the variational bound on negative log likehood $L$. The variational bound of $L$ is form by $\mathbf{x}_{0:m}$ and $\mathbf{y_r}$, which is given by
\begin{align}\label{var-bound}
  &L=\mathbb{E}\ [-\log\ p_{\mathbf{\theta}}(\mathbf{x}_0|\mathbf{h_r})]\le \mathbb{E}_q[-\log(\frac{p_{\mathbf{\theta}}(\mathbf{x}_{0:m},\mathbf{y_r}|\mathbf{h_r})}{q(\mathbf{x}_{1:m},\mathbf{y_r}|\mathbf{x}_0,\mathbf{h_r})})]\nonumber\\
  &=\mathbb{E}_q\ \underbrace{[D_{KL}(q(\mathbf{y_r}|\mathbf{x}_0,\mathbf{h_r})||p(\mathbf{y_r}|\mathbf{h_r}))}_{L_y}-\underbrace{\log p_{\mathbf{\theta}}(\mathbf{x}_0|\mathbf{x}_1,\mathbf{h_r})}_{L_0}\nonumber\\
  &+\underbrace{D_{KL}(q(\mathbf{x}_m|\mathbf{y_r},\mathbf{x}_0,\mathbf{h_r})||p_{\mathbf{\theta}}(\mathbf{x}_m|\mathbf{y_r},\mathbf{h_r}))}_{L_m}\nonumber\\
  &+{\sum_{t=1}^{m}\underbrace{D_{KL}(q(\mathbf{x}_{t-1}|\mathbf{x}_t,\mathbf{x}_0,\mathbf{h_r})||p_{\mathbf{\theta}}(\mathbf{x}_{t-1}|\mathbf{x}_t,\mathbf{h_r}))}_{L_{t-1}}}],
\end{align}
where $L_m$ instructs to select the hyperparameter $m$. In this paper, we select $m$ by
\begin{align}\label{selectm}
  arg\min_{m}\   2\sigma^{2}-\frac{1-\bar{\alpha}_m}{\bar{\alpha}_m}.
  %here can explain later
\end{align}

Similar to the process in \cite{Ho}, $L_{t-1}$  can be calculated in closed-form using the Rao-Blackwellized method. The optimization object of $L_{t-1}$ can be simplified by adopting re-parameterization and re-weighting methods as following
\begin{align}\label{simL}
  \mathbb{E}_{\mathbf{x}_0,\mathbf{\epsilon}}(||\mathbf{W_n\epsilon}-\mathbf{W_n\epsilon_{\theta}}(\mathbf{x}_t,\mathbf{h_r},t)||^2_2),
  %\mathbb{E}_{\mathbf{x_0,\epsilon}}(||\mathbf{W_n\epsilon}-\mathbf{W_n\epsilon_{\theta}}(\sqrt{\bar{\alpha_t}}\mathbf{x_0}+\sqrt{1-\bar{\alpha_t}}\mathbf{W_n}\epsilon)||^2_2)
\end{align}
\addtolength{\topmargin}{0.015in}
where $\mathbf{\epsilon_{\theta}}(\mathbf{x}_t,\mathbf{h_r},t)$ is the output of CDDM. Moreover, (\ref{simL}) can be re-weighted by ignoring the noise coefficient matrix $\mathbf{W_n}$ as following
\begin{align}\label{L_last}
\mathbb{E}_{\mathbf{x}_0,\mathbf{\epsilon}}(||\mathbf{\epsilon}-\mathbf{\epsilon_{\theta}}(\sqrt{\bar{\alpha}_t}\mathbf{x}_0+\sqrt{1-\bar{\alpha}_t}\mathbf{W_n}\mathbf{\epsilon})||^2_2).
\end{align}

Finally, to optimize (\ref{L_last}) for all $t\in\{1,2,...,T\}$, the loss function of the proposed CDDM is expressed as follows
\begin{align}\label{L_CDDM}
  L_{CDDM}(\mathbf{\theta})=\mathbb{E}_{\mathbf{x}_0,\mathbf{\epsilon},t}(||\mathbf{\epsilon}-\mathbf{\epsilon_{\theta}}(\sqrt{\bar{\alpha}_t}\mathbf{x}_0+\sqrt{1-\bar{\alpha}_t}\mathbf{W_n}\mathbf{\epsilon})||^2_2).
\end{align}

The training procedures of the proposed CDDM are summarized in Algorithm \ref{trainCDDM}.
%When training the model, $\mathbf{s}_1$ and $\mathbf{s}_2$ should be generated from the same dataset $S$ individually. A copy of the dataset $\tilde{S}$ is loaded with shuffling. $\mathbf{s}_1$ comes from the batch of $S$ and $\mathbf{s}_2$ comes from the batch $\tilde{S}$. For the FISF module, the SNR of the worse channel ($SNR_1$) is randomly set in a given range and we then randomly set $\gamma>0$ so that the SNR of better channel ($SNR_2$) is $SNR_1+\gamma$. Fusion rate $\alpha$ is also randomly selected in the range between 0 and 1 with step 0.1. When it is 0 or 1, which means only one source is expected to be delivered, the model degrades to an end-to-end model. The whole system takes $L$ as the loss function and all the parameters are updated jointly according to the loss $L$. The whole training procedures are described in Algorithm \ref{train}.
\begin{algorithm}[t]
  \hspace*{0.02in} {\bf \small{Input:}}
	\small{Training set $S$, hyper-parameter $T$ and $\bar{\alpha}_t$.} \\
	\hspace*{0.02in} {\bf \small{Output:}}
	\small{The trained CDDM.}
	\caption{Training algorithm of CDDM}
	\label{trainCDDM}
	\begin{algorithmic}[1] 
    \WHILE {the training stop condition is not met}
    \STATE Randomly sample $\mathbf{x}$ from $S$
    \STATE Randomly sample $t$ from $Uniform(\{1,...,T\})$
    \STATE Sapmle $|\mathbf{h_c}|$ and compute $\mathbf{H_r}$, $\mathbf{W_s}$ and $\mathbf{W_n}$
    \STATE Randomly sample $\mathbf{\epsilon}$ from $\mathcal{N}(0,\mathbf{I}_{2k})$
    \STATE Take gradient descent step according to (\ref{x_t}) and (\ref{L_CDDM})\\
    $\nabla_\mathbf{\theta}(||\mathbf{\epsilon}-\mathbf{\epsilon _\theta}(\sqrt{\bar{\alpha}_t}\mathbf{W_sx}+\sqrt{1-\bar{\alpha}_t}\mathbf{W_n}\mathbf{\epsilon})||^2_2)$
    %\IF {$\alpha==0$}
    %\STATE Set $\beta=0$ and then compute loss $L=L_2$.
    %\ELSIF {$\alpha==1$}
    %\STATE Compute loss $L=L_1$.
    %\ELSE
    %\STATE Compute loss $L=L_1+\lambda L_2$.
    %\ENDIF
    %\STATE Update all the parameters to minimize $L$.
    \ENDWHILE
	\end{algorithmic}
\end{algorithm}

\subsection{Sampling Algorithm of CDDM}\label{B}
To reduce the time consumption of sampling process, (\ref{var-bound}) implies that selecting $m$ according to (\ref{selectm}) and setting $\mathbf{x}_m=\mathbf{y_r}$ is a promising way. By utilizing the received signal $\mathbf{y_r}$, only $m$ steps are needed to be excuted. For each time step $t\in\{1,2,...,m\}$, the trained CDDM outputs $\mathbf{\epsilon_\theta}(\mathbf{x}_t,\mathbf{h_r},t)$, which attempts to predict $\mathbf{\epsilon}$ from $\mathbf{x}_t$ without knowledge of $\mathbf{x}_0$. A sampling algorithm is required to sample $\mathbf{x}_{t-1}$. The process is excuted for $m$ times such that $\mathbf{x}_0$ can be computed out finally.

We first define the sampling process $f(\mathbf{x}_{t-1})$ with the knowledge of $\mathbf{\epsilon}$ as following
\begin{align}
  f(\mathbf{x}_{t-1})=q(\mathbf{x}_{t-1}|\mathbf{x}_t,\mathbf{x}_0,\mathbf{h_r}).
\end{align}

Applying Bayes rule, the distribution can be expressed as a Gaussian distribution
\begin{gather}
  q(\mathbf{x}_{t-1}|\mathbf{x}_t,\mathbf{x}_0,\mathbf{h_r})\nonumber\\
  \sim \mathcal{N}(\mathbf{x}_{t-1};\sqrt{\bar{\alpha}_{t-1}}\mathbf{x}_0+\sqrt{1-\bar{\alpha}_{t-1}}\frac{\mathbf{x}_t-\sqrt{\bar{\alpha}_t}\mathbf{x}_0}{\sqrt{1-\bar{\alpha}_t}},0),
\end{gather}
where $\mathbf{x}_0$ is acquired by re-writing (\ref{reforward}) as following
\begin{align}
  \mathbf{x}_0=\frac{1}{\sqrt{\bar{\alpha}_t}}(\mathbf{x}_t-\sqrt{1-\bar{\alpha}_t}\mathbf{W_n}\mathbf{\epsilon}).
\end{align}

However, only $\mathbf{\epsilon_\theta}(\mathbf{x}_t,\mathbf{h_r},t)$ is available for sampling. $\mathbf{x}_0$ is derived through an estimation process by replacing $\mathbf{\epsilon}$ with $\mathbf{\epsilon_\theta}(\mathbf{x}_t,\mathbf{h_r},t)$ as following
\begin{align}\label{predictx-0}
  {\hat{\mathbf{x}}_0}=\frac{1}{\sqrt{\bar{\alpha}_t}}(\mathbf{x}_t-\sqrt{1-\bar{\alpha}_t}\mathbf{W_n}\mathbf{\epsilon_\theta}(\mathbf{x}_t,\mathbf{h_r},t)).
\end{align}

As a result, the sampling process is replaced with
\begin{align}
  f_\mathbf{\theta}(\mathbf{x}_{t-1})=p_\mathbf{\theta}(\mathbf{x}_{t-1}|\mathbf{x}_t,\hat{\mathbf{x}}_0,\mathbf{h_r}).
\end{align}

Without the knowledge of $\mathbf{\epsilon}$, a sample of $\mathbf{x}_{t-1}$ is
\begin{align}e
  \mathbf{x}_{t-1}=&\sqrt{\bar{\alpha}_{t-1}}\underbrace{(\frac{1}{\sqrt{\bar{\alpha}_t}}(\mathbf{x}_t-\sqrt{1-\bar{\alpha}_t}\mathbf{W_n}\mathbf{\epsilon_\theta}(\mathbf{x}_t,\mathbf{h_r},t)))}_{estimate\ \mathbf{x}_0}\nonumber\\
  &+\underbrace{\sqrt{1-\bar{\alpha}_{t-1}}\mathbf{W_n\epsilon_\theta}(\mathbf{x}_t,\mathbf{h_r},t)}_{sample\ \mathbf{x}_{t-1}}.
\end{align}

Note that for the last step $t=1$, we only predict $\mathbf{x_0}$ such that sampling is taken as
\begin{align}
  \mathbf{y}=\frac{1}{\sqrt{\bar{\alpha}_1}}(\mathbf{x}_1-\sqrt{1-\bar{\alpha}_1}\mathbf{W_n\epsilon_\theta}(\mathbf{x}_1,\mathbf{h_r},1)).
\end{align}
The sampling method is summarized in Algorithm \ref{sampleCDDM}.

\begin{algorithm}[t]
  \hspace*{0.02in} {\bf \small{Input:}}
  \small{$\mathbf{y_r}$,$\mathbf{h_r}$,hyperparameter $m$} \\
  \hspace*{0.02in} {\bf \small{Output:}}
  \small{$\mathbf{y}$}
  \caption{Sampling algorithm of CDDM}
  \label{sampleCDDM}
  \begin{algorithmic}[1]
    \STATE $\mathbf{x}_m=\mathbf{y_r}$ 
    \FOR {$t=m,...,2$}
    \STATE $\mathbf{z}=\mathbf{W_n}\mathbf{\epsilon_\theta}(\mathbf{x}_t,\mathbf{h_r},t)$
    \STATE $\mathbf{x}_{t-1}=\sqrt{\bar{\alpha}_{t-1}}(\frac{\mathbf{x}_t-\sqrt{1-\bar{\alpha}_t}\mathbf{z}}{\sqrt{\bar{\alpha}_t}})+\sqrt{1-\bar{\alpha}_{t-1}}\mathbf{z}$
    \ENDFOR
    \STATE $t=1$
\STATE $\mathbf{z}=\mathbf{W_n}\mathbf{\epsilon_\theta}(\mathbf{x}_1,\mathbf{h_r},1)$
    \STATE $\mathbf{y}=\frac{\mathbf{x}_1-\sqrt{1-\bar{\alpha}_{1}}\mathbf{z}}{\sqrt{\bar{\alpha}_1}}$.
  \end{algorithmic}
\end{algorithm}

\section{Application of CDDM in Semantic Communications System Based on JSCC}
In this section, the proposed CDDM is applied into a semantic communications system based on JSCC for wireless image transmission.
\subsection{System Structure}
An overview architecture of the joint CDDM and JSCC system is shown in Fig. \ref{CDESC}. An RGB source image $\mathbf{s}$ is encoded by a JSCC encoder. In this paper, the JSCC is built upon the Swin Transformer\cite{Ze} backbone, which has a more powerful expression ability than vision transformer by replacing the standard multi-head self attention in vision transformer with a shift window multi-head self attention. Two convolution layers are adopted as the output layer of the JSCC encoder, constituting variational auto-encoder (VAE)\cite{Diederik} structure. The JSCC encoder computes the source image $\mathbf{s}$ as $\mathbf{\mu _\phi}\in \mathbb{R}^{2k}$ and $\mathbf{\sigma_\phi}\in \mathbb{R}^{2k}$. Finally, the JSCC encoder samples the transmitted signal $\mathbf{x}$ as
\begin{align}
  \mathbf{x}=\mathbf{\mu_\phi+\sigma_\phi\xi},
\end{align}
where $\mathbf{\phi}$ encapsulates all parameters of the JSCC encoder and $\mathbf{\xi}\sim \mathcal{N}(0,\mathbf{I}_{2k})$. $\mathbf{x}$ is then tranmitted and processed into $\mathbf{y_r}$ at the receiver, as described in Section \ref{I}. At the receiver, the proposed CDDM removes the channal noise from $\mathbf{y_r}$ using Algorithm \ref{sampleCDDM}. Following this, the output of CDDM is fed into the JSCC decoder to reconstruct the source image $\mathbf{\hat{s}}$.

\subsection{Training algorithm}
\addtolength{\topmargin}{0.015in}
The entire training algorithm of the joint CDDM and JSCC system consists of three stages. In the first stage, the JSCC encoder and decoder are trained jointly through the channel shown in Fig. \ref{CDESC}, except for the CDDM module, to minimize the distance $d(\mathbf{s,\hat{s}})$. MSE is used as the performance metric and a slight KL divergence punishment with normal distribution is exerted on the JSCC encoder. The slight punishment does not reduce the final performance but it can constraint $\mathbf{x}$ in a more structured way, thereby enhancing the convergence of CDDM in the second stage. Therefore, the loss function for this stage is given by
\begin{align}
  L_1(\mathbf{\phi,\varphi})&=\mathbb{E}_{\mathbf{s}\sim p_\mathbf{s}}\mathbb{E}_{\mathbf{y_r}\sim p_{\mathbf{y_r|s}}}||\mathbf{s}-\mathbf{\hat{s}}||^2_2\nonumber\\
  &+\lambda D_{KL}(p(\mathbf{x|s})||N(0,\mathbf{I}_{2k})),
\end{align}
where $\mathbf{\varphi}$ encapsulate all parameters of JSCC decoder and $\lambda$ is the punishment weight.

In the second stage, the parameters of the JSCC encoder are fixed such that CDDM can learn the distribution of $\mathbf{x}_0$ via Algorithm \ref{trainCDDM}. The training process is not affected by the channel noise power because Algorithm \ref{trainCDDM} has a special noise schedule, and the noise has been designed specially to simulate the distribution of channel noise. Benefitting from this, CDDM is designed for handling various channel conditions and requires only one training process.

In the third stage, the JSCC decoder is re-trained jointly with the trained JSCC encoder and CDDM to minimize $d(\mathbf{s,\hat{s}})$. The entire joint CDDM and JSCC system is performed through the real channel, while only the parameters of the decoder are updated. The loss function is derived as
\begin{align}
  L_3(\mathbf{\varphi})=\mathbb{E}_{\mathbf{y}\sim p_{\mathbf{y|s}}}||\mathbf{s}-\mathbf{\hat{s}}||^2_2.
\end{align}
The training algorithm is summarized in Algorithm \ref{trainCDESC}.

\section{EXPERIMENTS RESULTS}
\begin{algorithm}[t]
  \hspace*{0.02in} {\bf \small{Input:}}
	\small{Training set $p(\mathbf{s})$, hyper-parameter $T$, $\bar{\alpha_t}$, and the channel estimation result $\mathbf{h_c}$ and $\sigma^2$.} \\
	\hspace*{0.02in} {\bf \small{Output:}}
	\small{The trained joint CDDM and JSCC system.}
	\caption{Training algorithm of the joint CDDM and JSCC system}
	\label{trainCDESC}
	\begin{algorithmic}[1]
    \WHILE {the training stop condition of stage one is not met }
    \STATE Randomly sample $\mathbf{s}$ from $S$
    \STATE Perform forward propagation through channel without CDDM.
    \STATE Compute $L_1(\mathbf{\mathbf{\phi, \varphi}})$ and update $\mathbf{\phi, \varphi}$
    \ENDWHILE
    \WHILE {the training stop condition of stage two is not met }
    \STATE Randomly sample $\mathbf{s}$ from $S$
    \STATE Compute $\mathbf{s}$ as $\mathbf{x}$
    \STATE Train CDDM with Algorithm \ref{trainCDDM}.
    \ENDWHILE
    \WHILE {the training stop condition of stage three is not met }
    \STATE Randomly sample $\mathbf{s}$ from $S$
    \STATE Perform forward propagation through channel with noise power $\sigma^2$ with the trained CDDM
    \STATE Compute $L_3(\mathbf{\varphi})$ and update $\mathbf{\varphi}$
    \ENDWHILE
	\end{algorithmic}
\end{algorithm}

In this section, we provide experiments results to verify the effectiveness of the proposed CDDM. In the experiments, the CDDM is established on U-Net architecture similar to \cite{Ho}, which accommodates $\mathbf{x}$ and $\mathbf{h_r}$ as input components. We use CIFAR10\cite{Krizhevsky} dataset for training and testing. We set $T=1000$ and $\lambda=5\times 10^{-5}$. We set $\alpha_t$ to constants decreasing linearly from $\alpha_1=0.9999$ to $\alpha_T=0.98$.

We adopt the JSCC system and classical separation-based source and channel coding scheme as the benchmarks for our performance comparison. It should be noted that in both the joint CDDM and JSCC system, as well as the JSCC system, we have used the same structure for JSCC. For the JSCC system, each SNR requires its corresponding model to be trained. The channel bandwith ratio is set as $\frac{1}{8}$. 
For the classical scheme, we employ the JPEG2000 codec for compression and LDPC\cite{DVB} codec for channel coding, marking as ``JPEG2000+LDPC".

\begin{figure}[t]
\begin{center}
 \includegraphics[width=7.6cm]{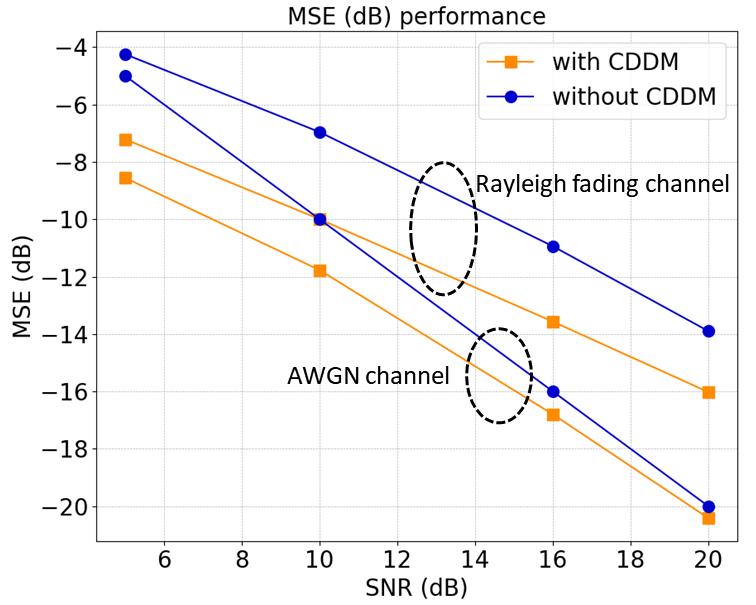}
\end{center}
 \caption{The MSE performance of the proposed CDDM versus SNR over different channels.}
\label{MSE}
\end{figure}

Fig. \ref{MSE} illustrates the MSE performance of CDDM in different signal-to-noise ratio (SNR) regimes. In the case of using CDDM, we caculate the MSE between $\mathbf{x}$ and $\mathbf{y}$, while in the case of not using CDDM, we calculate the MSE between $\mathbf{x}$ and $\mathbf{y}_r$. As shown in Fig. 1, $\mathbf{y}_r$ and $\mathbf{y}$ are the input and output of CDDM, respectively. We can see that the system with CDDM performs much better than the system without CDDM in all SNR regimes under both AWGN and Rayleigh fading channels. For example, for AWGN channel, the proposed CDDM has a $0.49$ dB gain in MSE at SNR=$20$ dB. Meanwhile, it can be seen that as the SNR decreases, the gain of CDDM in MSE increases. This indicates that as the SNR decreases, i.e., the channel noise increases, the proposed CDDM is easier to remove more noise, e.g. $3.55$ dB gain at SNR=$5$ dB for AWGN channel. Moreover, it is important to note that under Rayleigh fading channel, MMSE has theoretically minimized the MSE, but CDDM can further reduce the MSE after MMSE. The reason for this fact is that CDDM can learn the distribution of $ \mathbf{x}_0=\mathbf{W_sx}$, and utilizes this learned knowledge to remove the noise, improving the effective SNR and thereby further reducing the MSE.

\begin{figure}[t]
\begin{center}
 \includegraphics[width=7.6cm]{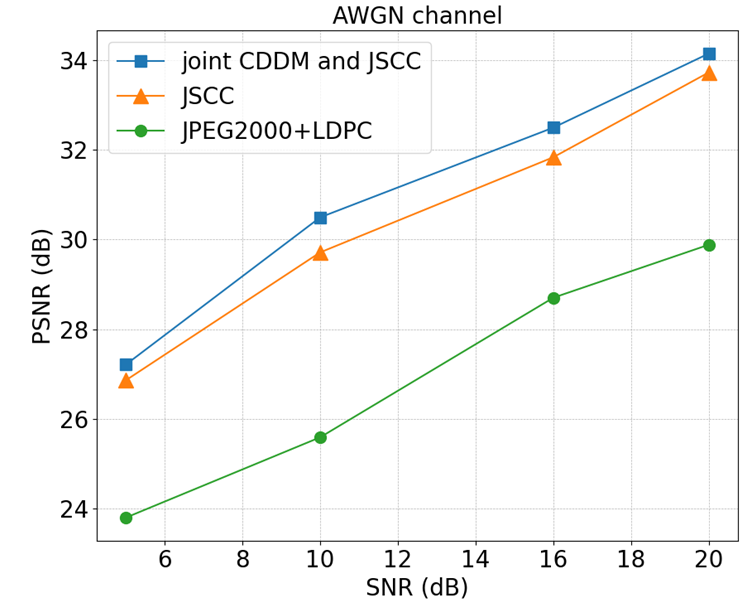}
\end{center}
 \caption{The PSNR performance versus SNR over AWGN channel.}
\label{PSNR-AWGN}
\end{figure}

\begin{figure}[t]
\begin{center}
 \includegraphics[width=7.6cm]{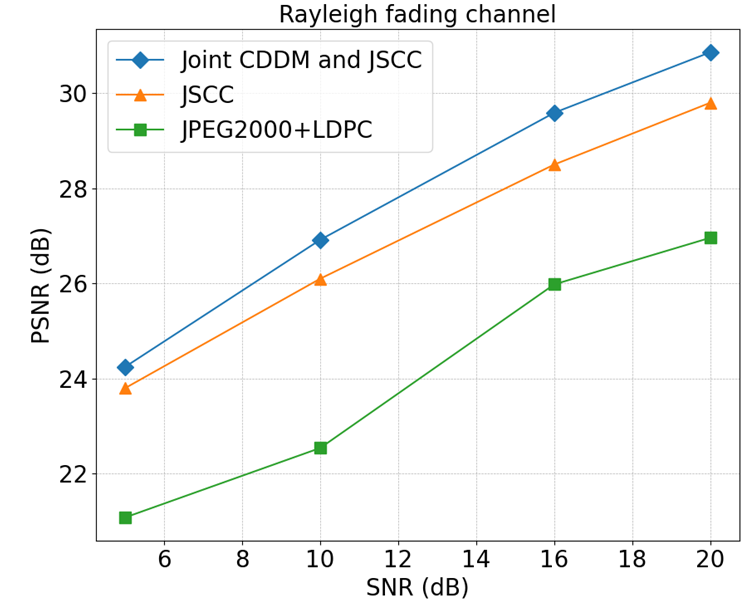}
\end{center}
 \caption{The PSNR performance versus SNR over Rayleigh fading channel.}
\label{PSNR-rayleigh}
\end{figure}

Fig. \ref{PSNR-AWGN} and Fig. \ref{PSNR-rayleigh} show the PSNR performance versus SNR under AWGN channel and Rayleigh fading channel, respectively. Given a SNR, both the joint CDDM and JSCC system and the JSCC system need to be retrained to achieve the best performance. Under both Rayleigh fading and AWGN channels, the joint CDDM and JSCC system achieves better PSNR performance compared to the JSCC system at  SNR ranging from $5$ dB to $20$ dB. For example, compared to the JSCC system, the joint CDDM and JSCC system achieves $1.06$ dB gain at SNR=$20$ dB over Rayleigh fading channel. Moreover, we also can observe that the CDDM and JSCC system significantly outperforms the ``JPEG2000+LDPC" scheme over both Rayleigh fading and AWGN channels.

\section{CONCLUSION}
In this paper, we have proposed the channel denoising diffusion models to eliminate the channel nosie under Rayleigh fading channel and AWGN channel. CDDM is trained utilizing a specialized noise schedule adapted to the wireless channel, which permits effective elimination of the channel noise via a suitable sampling algorithm in the reverse sampling process. CDDM is then applied into the semantic communications system based on JSCC. Experimental results show that under both AWGN and Rayleigh fading channels, the system with CDDM performs much better than the system without CDDM in terms of MSE and PSNR. 

\footnotesize
\bibliographystyle{IEEEtran}
\bibliography{reference}{}
\end{document}